\pdfoutput=1
\documentclass[twoside,twocolumn,a4paper,10pt]{article}

\usepackage[utf8]{inputenc}
\usepackage[T1]{fontenc}
\usepackage{mathptmx}
\usepackage[scaled=0.92]{helvet}
\usepackage{courier}
\usepackage{amsmath}
\usepackage{booktabs}
\usepackage[normalem]{ulem}
\usepackage{float}
\usepackage{graphicx}
\usepackage{capt-of}
\usepackage{lastpage}
\usepackage{cite}

\makeatletter

\ifdefined\pdfpagewidth \pdfpagewidth=\paperwidth \pdfpageheight=\paperheight \fi

\renewcommand\normalsize{\@setfontsize\normalsize{9}{11}%
  \abovedisplayskip 5.5pt plus 2pt minus .5pt
  \belowdisplayskip \abovedisplayskip
  \abovedisplayshortskip 5.5pt plus 2pt
  \belowdisplayshortskip \abovedisplayshortskip
  \let\@listi\@listI}
\renewcommand\small{\@setfontsize\small{8}{10}%
  \abovedisplayskip 5pt plus 1pt minus .5pt
  \belowdisplayskip \abovedisplayskip
  \abovedisplayshortskip 5pt plus 1pt
  \belowdisplayshortskip \abovedisplayshortskip
  \def\@listi{\leftmargin\leftmargini \topsep\z@ plus 1pt minus .5pt
    \parsep\z@ \itemsep\z@}}
\renewcommand\footnotesize{\@setfontsize\footnotesize{8}{10}}
\renewcommand\scriptsize{\@setfontsize\scriptsize{6}{7}}
\renewcommand\tiny{\@setfontsize\tiny{6}{8}}
\renewcommand\large{\@setfontsize\large{10}{11}}
\renewcommand\Large{\@setfontsize\Large{12}{14}}
\renewcommand\LARGE{\@setfontsize\LARGE{14}{17}}
\renewcommand\huge{\@setfontsize\huge{17}{20}}
\renewcommand\Huge{\@setfontsize\Huge{20}{25}}

\def\@listI{\leftmargin\leftmargini \parsep\z@ \partopsep\z@
  \topsep\z@ plus 1pt minus .5pt \itemsep\z@ plus .1pt}
\let\@listi\@listI
\normalsize

\long\def\@makecaption#1#2{\vskip 6pt
  \setbox\@tempboxa\hbox{\textbf{#1:} \normalfont\textit{#2}}%
  \ifdim\wd\@tempboxa>\hsize
    \textbf{#1:} \normalfont\itshape #2\par
  \else
    \hbox to\hsize{\hfil\box\@tempboxa\hfil}%
  \fi
  \vskip 6pt}

\renewcommand\section{\@startsection{section}{1}{\z@}%
  {-11pt plus -5.5pt minus -1pt}{0.001pt}%
  {\raggedright\normalfont\normalsize\bfseries}}
\renewcommand\subsection{\@startsection{subsection}{2}{\z@}%
  {-11pt plus -5.5pt minus -1pt}{0.001pt}%
  {\raggedright\normalfont\normalsize\bfseries}}
\renewcommand\subsubsection{\@startsection{subsubsection}{3}{\z@}%
  {-5.5pt plus -5.5pt minus -1pt}{0.001pt}%
  {\raggedright\normalfont\normalsize\bfseries}}
\renewcommand\paragraph{\@startsection{paragraph}{4}{\z@}%
  {5.5pt plus 2.25pt minus 1pt}{-0.5em}%
  {\normalfont\normalsize\bfseries}}
\renewcommand\@seccntformat[1]{\csname the#1\endcsname.\hskip 0.5em}

  {\endlist}
\def\@biblabel#1{[#1] \ }

\newcommand\paper@shortauthor{Do\u{g}an, Ba\c{s}aran, Sezgin, Montazeri}
\newcommand\paper@shorttitle{From Sketch to Scene}
\newcommand\paper@pagenum{\thepage\ of \pageref*{LastPage}}

\def\ps@preprinttitle{\let\@mkboth\@gobbletwo
  \let\@oddhead\@empty \let\@evenhead\@empty
  \let\@oddfoot\@empty \let\@evenfoot\@empty}
\def\ps@preprint{\let\@mkboth\@gobbletwo
  \def\@oddhead{\small\hfill\itshape\paper@shortauthor\ / \paper@shorttitle
    \hfill\upshape\paper@pagenum}%
  \def\@evenhead{\small\upshape\paper@pagenum\hfill
    \itshape\paper@shortauthor\ / \paper@shorttitle\hfill}%
  \let\@oddfoot\@empty \let\@evenfoot\@empty}

\newcommand\paperrule{\par{\parskip\z@\noindent\rule{\textwidth}{0.25pt}\par}}

\makeatother

\usepackage[pagebackref=true,colorlinks,linkcolor=blue,citecolor=blue,
  urlcolor=blue,bookmarks=false,pdfpagemode=UseNone]{hyperref}
\providecommand\webLink[1]{\url{#1}}
\hypersetup{pdftitle={From Sketch to Scene},
  pdfauthor={Doğan, Başaran, Sezgin, Montazeri}}

\begin{document}

\twocolumn[{%
  \vspace*{12pt}
  {\centering
    {\huge\bfseries From Sketch to Scene: Relightable Garment Reconstruction from Marked Sketches\par}
    \vskip 24pt
    {\normalsize
    \begin{tabular}[t]{@{}c@{}}
    \parbox{\textwidth}{\centering
    Ahmet Neçirvan Doğan$^{1,*}$ \qquad
    Eren Acar Başaran$^{1,*}$ \qquad
    Metin Sezgin$^{1}$ \qquad
    Zahra Montazeri$^{2}$
    \\[8pt]
    $^{1}$ Koç University, Türkiye
    \\
    $^{2}$ University of Manchester, UK
    \\[8pt]
    {\small $^{*}$ Equal contribution}
    }
    \end{tabular}\par}
    \vskip 24pt
  }
  \includegraphics[width=1.0\linewidth]{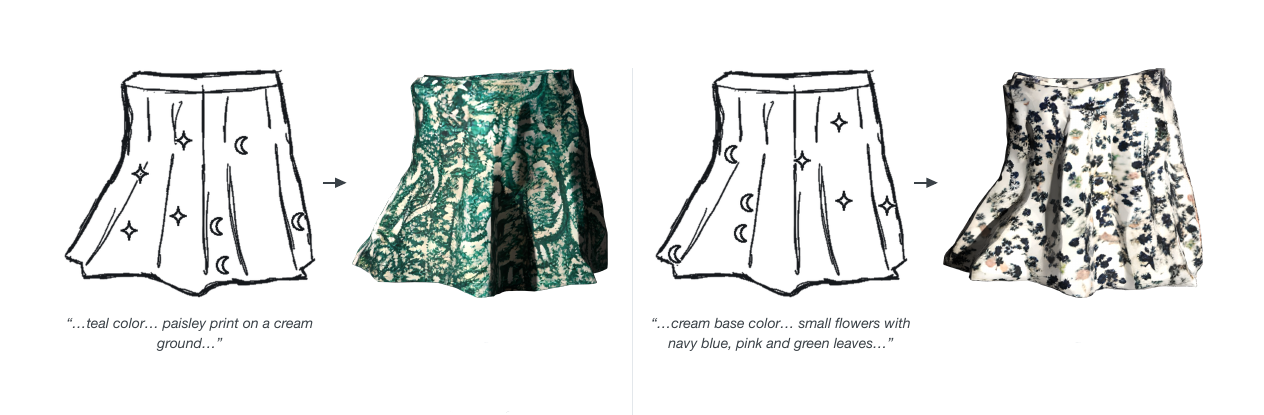}
  \captionof{figure}{\textbf{From a marked sketch to a relightable garment}
  A hand-drawn sketch with star marks for highlights and moon marks for shadows,
  together with a texture prompt, is turned into a textured 3D garment rendered
  under the directional light predicted from its marks.
  \emph{Left:} one placement of the marks and one prompt.
  \emph{Right:} the same sketch with the marks moved and a different prompt.}
  \label{fig:teaser}
  \paperrule
  \vspace{2pt}
  \begin{list}{}{\leftmargin 2pc\rightmargin\leftmargin\parsep 0pt plus .1pt}
  \item[]{\textbf{Abstract}}\\\itshape
Sketching is one of the most natural ways for people to express visual ideas. Unlike the text prompts that most generative systems rely on today, a sketch captures spatial structure and intent directly, including where a designer wants highlights and shadows to appear. Yet current sketch-to-3D pipelines have no input for that intent: they bake the drawn shading into the texture as dark paint, and the light cannot be moved. We propose a system that transforms a rough garment sketch, marked with sparse annotations---a star for a highlight and a moon for a shadow---and a short text prompt describing the texture, into a textured 3D mesh with an explicit directional light, as a scene that can be relit and edited after generation. The two symbols are the most frequent proposals in an elicitation study with art students. From the marked sketch the system predicts the scene's directional light and, separately, a displacement field that bends the surface so that the moon-marked regions fall into shadow under that light, rather than being painted dark in the texture.

  \vspace{3pt}
  {\upshape\bfseries CCS Concepts}\par
  \textbullet\ \textbf{Computing methodologies} $\,\to\,$ \textbf{Shape modeling; } \textbf{Reconstruction; }
  \textbullet\ \textbf{Human-centered computing} $\,\to\,$ \textbf{Interaction techniques; }
  \end{list}
  \paperrule
  \vspace{14pt}
}]
\thispagestyle{preprinttitle}

  \section{Introduction}
  \label{sec:introduction}

Fashion design begins on paper. A designer sketches a garment's
silhouette, then adds how the fabric catches the light: its folds,
its shadows and its highlights \cite{riegelman2012nineheads}. Today,
a sketch can be turned into a textured 3D model without a 3D artist:
an image generator produces a plausible rendering from it
\cite{zhang2023controlnet}, and a reconstruction method lifts that
rendering into a mesh \cite{xiang2025structured,hunyuan2025hunyuan3d21}.
Such pipelines have no input for the designer's shading intent:
shading strokes are either treated as line art or reproduced as
appearance.

The result is that a designer can specify what a garment looks like
but not how it is lit. Text prompts do not help much here. A phrase such as ``a soft shadow under the left sleeve'' describes a location, but a text-conditioned generator has no explicit way to tie the phrase to a region of the drawing, and follows it only loosely.

We propose treating shading marks as a first-class input. A
designer marks the sketch with a two-symbol notation elicited from art
students: a star where a highlight should fall, and a moon where the
fabric folds into shadow. The marks are placed on the drawing itself. Our pipeline reads them
and produces a relightable 3D garment whose lighting
and surface both follow where the marks were placed
(Figure~\ref{fig:teaser}).

In summary, we make the following contributions:
\begin{itemize}
\item We conduct an elicitation study with art students and derive
from it a two-symbol shading notation: a star for a highlight and a
moon for a shadow (Secs.~\ref{sec:elicitation}
and~\ref{sec:representation}).

\item We predict the scene's directional light from these marks, so
the light is an editable part of the scene rather than fixed in the
texture (Secs.~\ref{sec:irradiance} and~\ref{sec:light}).

\item We predict a displacement field from the marks and use it to
displace the generated mesh, so that the marked regions fall into
shadow under the predicted light
(Secs.~\ref{sec:field},~\ref{sec:displacement} and~\ref{sec:shading}).

\item We combine these into a pipeline that turns a marked sketch and
a texture prompt into a relightable 3D scene
(Sec.~\ref{sec:method-overview}).
\end{itemize}

  \section{Related work}
  \label{sec:related-work}
Our input is a sketch with shading marks and a prompt specifying the texture, and our output is a 3D
garment with an explicit directional light. We therefore review sketch-based modeling
(Sec.~\ref{sec:related-sketch-modeling}), how artists and 3D tools go
about specifying lighting (Sec.~\ref{sec:related-specifying-lighting}),
and the text- and image-conditioned 3D generation our pipeline builds on
(Sec.~\ref{sec:related-text-image-3d}).

  \subsection{Sketch-based modeling}
  \label{sec:related-sketch-modeling}

  Early interactive systems read strokes as geometric operations: Teddy
\cite{igarashi1999teddy} inflates a closed contour into a freeform
surface, and FiberMesh \cite{nealen2007fibermesh} keeps the drawn curves
as editable handles on the result. Structured annotations
\cite{gingold2009structured} let the user place primitives and
constraints, such as symmetry and equal length, on a 2D drawing to settle
its 3D interpretation. Garments have their own line of work, because a
drawn garment is defined relative to a body. Turquin et
al.~\cite{turquin2007sketch} lift contours drawn over a view of a 3D body
model into a garment around it, and FoldSketch \cite{li2018foldsketch}
lets a designer draw fold strokes on an existing garment and solves for a
sewing pattern that physically reproduces them.

Learned methods replace these rules with priors taken from data. From a single drawing, networks predict multi-view depth and normal maps
\cite{lun2017multiview,li2018robust} or infer shape and viewpoint jointly
\cite{zhang2021sketch2model}. For clothing, Wang et
al.~\cite{wang2018learning} let the user sketch the fold patterns they
want and recover the garment and body-shape parameters that produce
them. Diffusion models have since made the sketch one condition among
several \cite{zheng2023lasdiffusion,binninger2024sens}. A common route
today passes the sketch through a conditioned image generator
\cite{zhang2023controlnet} and lifts the result with a single-image 3D
generator (Sec.~\ref{sec:related-text-image-3d}).

A related thread reads shading \emph{out of} line drawings \cite{hudon2018deep}. Lumo
\cite{johnston2002lumo} interpolates normals from contours to light cel
animation; CrossShade \cite{shao2012crossshade} infers surface normals
from the cross-section curves of concept sketches; Ink-and-Ray
\cite{sykora2014inkandray} builds bas-relief proxies so that hand-drawn
characters receive global illumination; and Zheng et
al.~\cite{zheng2020learning} add shadows to line art given a light
direction. They recover normals, shading or a proxy shape, and leave the light
for the user to supply; none reads it from the drawing.

In all of this work the sketch describes geometry. Shading a designer
draws is read as more line work, or, in data-driven pipelines, comes
back as texture on the result. To our knowledge, no existing method
reads a drawing for where the designer intends light to fall and
returns that intent as a light in the scene.

  \subsection{Specifying lighting}
  \label{sec:related-specifying-lighting}

Our marks specify lighting on a drawing, so we look at how light is
handled in two settings: how artists draw it, by showing where it falls
and where shadows form, and how 3D tools set it, as an object the user
can move and edit.

\paragraph{How artists draw light} Artists usually draw light through what it does to a form. Painting manuals teach a small vocabulary of
shading events on a form: the highlight, the terminator where the surface
turns away from the light, the core shadow beyond it, and the cast shadow
\cite{gurney2010color}. Non-photorealistic rendering reproduces these
conventions from 3D, placing hatching according to a known light
\cite{hertzmann2000illustrating}; there the light is an input. Our
elicitation (Sec.~\ref{sec:elicitation}) asks a different question:
which symbols people would use to indicate where light and shadow fall.

\paragraph{How 3D tools set light} 3D tools instead expose the light source itself. Blender, for example, exposes light as an explicit
object, such as a point, spot, area or directional (sun) source, whose
orientation, color and intensity the artist edits in the viewport
\cite{blender2024manual}, and glTF carries the same parameters between
tools \cite{khronos2018lights}. Light is then editable after the fact, but
the user must reason about causes: to put a highlight in a chosen place,
the artist moves the source and watches the result. Some research methods remove this step: the user specifies the effect,
such as a highlight or a painted target, and the method solves for the light \cite{okabe2007illumination,pellacini2010envylight}. Poulin and Fournier \cite{poulin1992lights}
recover light positions from highlights and shadows sketched onto a
rendered scene, and Pellacini et al.~\cite{pellacini2007lighting} solve
for light parameters that reproduce a painted target illumination;
Schmidt et al.~\cite{schmidt2016state} survey the wider space. Our marks
are closest to Poulin and Fournier's highlight and shadow sketches. These
methods, however, operate on an existing 3D scene, where each candidate
light can be rendered and compared with the user's target. In our setting there is no scene yet, only a drawing, so the light has to be inferred before any geometry exists.

  \subsection{Text- and image-conditioned 3D generation}
  \label{sec:related-text-image-3d}
Current 3D generators work feed-forward. Novel-view diffusion and large
reconstruction models lift a single image to a mesh in seconds
\cite{liu2023zero1to3,hong2024lrm}, and native 3D latent generators such
as TRELLIS \cite{xiang2025structured} and Hunyuan3D
\cite{hunyuan2025hunyuan3d21} produce high-quality geometry, textured by
a separate multi-view painting stage. Commercial services such as Meshy
\cite{meshy2025} put the same pipeline behind a text or image prompt.

For our purposes these systems share two properties. First, their
control channels are text and images, so shading can only be
specified by handing them a picture that already contains it. Second, they return geometry and material but no light.
Recent texture stages output physically based materials instead of a single color map \cite{hunyuan2025hunyuan3d21}, so their assets can be
relit in principle. However, the illumination of the conditioning image
is treated as a nuisance and removed from the albedo
\cite{zeng2024paint3d}; where removal is incomplete, it stays in the
texture. Either way, the output carries no light of its own, and any lighting
left in the texture cannot be moved. We use such a generator to build the mesh and paint its texture, and
add the light as a separate element of the scene.

  \section{Shading marks}
  \label{sec:shading-marks}
This section introduces the shading marks, the input that separates our
system from other sketch-to-3D pipelines. We first set out what the
notation must achieve (Sec.~\ref{sec:shading-marks-motivation}), then
describe the study we used to choose the symbols
(Sec.~\ref{sec:elicitation}), and finally define what each symbol means
and how it is placed (Sec.~\ref{sec:representation}).

  \subsection{Motivation}
  \label{sec:shading-marks-motivation}
Our goal is to let a designer say, directly on the sketch, how the
garment should be lit: where light falls and where the fabric should
be in shadow. For such marks to be useful, they need two
properties. A designer without training in shading should be able to
make them and to guess what each one means without being taught. Both
depend on the symbols themselves, so instead of inventing symbols
we asked people which ones they would use (Sec.~\ref{sec:elicitation}).

  \subsection{Elicitation}
  \label{sec:elicitation}

To choose the symbols, we asked art students which symbols they would
use to mark a highlight and a shadow. We first describe the
questionnaire and how the drawn answers were coded, then report which
symbols were proposed most often and how the two categories differed.

\paragraph{Participants and procedure} We ran an elicitation study
\cite{wobbrock2005maximizing,wobbrock2009user} with 28 art students at our university in April 2026. The study was introduced in an information session during class, and participation was
voluntary. Participants completed a paper questionnaire: two background
items (drawing experience and formal training in light and shading), a
free drawing task, two items on how they had depicted highlights and
shadows, and the target item:
\emph{``If you had to encode shading using ONLY symbols, what would you
choose? Fill one symbol per category: Highlight: \_\_\_ , Shadow: \_\_\_''}.
Answers were free-form drawings. Most participants were novices: 18
described themselves as beginners with no formal training, 8 as
intermediate and 2 as advanced, and 5 of 28 had studied light and shading
formally.

\paragraph{Coding} Each drawn answer was transcribed to the nearest
symbol class. One rule separates a symbol from a technique: hatching,
meaning multiple parallel strokes, describes \emph{how} to shade rather
than naming a symbol, and is excluded as off-task, while a single stroke
counts as a mark. The rule is applied identically to both categories.
Blank answers are non-responses. This leaves 25 highlight and 22 shadow
proposals (Table~\ref{tab:elicitation}). We report the share of the modal
proposal with a Wilson 95\% interval \cite{wilson1927}, the agreement rate $A$
\cite{wobbrock2005maximizing}, and its bias-corrected form $AR$, for which
Vatavu and Wobbrock define magnitude bands \cite{vatavu2015formalizing}.

\begin{table}[t]
  \centering
  \caption{Symbols proposed for highlight and shadow by 28 participants.
  Hatching is a shading technique instead of a symbol and is excluded
  before agreement is computed. ``Other'' groups proposals made by a
  single participant each.}
  \label{tab:elicitation}
  \small
  \begin{tabular}{lrr}
    \toprule
    Proposal & Highlight & Shadow \\
    \midrule
    Star / sparkle        & \textbf{14} & --        \\
    Moon              & --          & \textbf{6} \\
    Circle                & 3           & 4         \\
    Line / dash           & 2           & 2         \\
    Spiral                & --          & 2         \\
    Other (one each)      & 6           & 8         \\
    \midrule
    Valid proposals       & 25          & 22        \\
    Hatching (excluded)   & 0           & 6         \\
    Blank                 & 3           & 0         \\
    \midrule
    Modal share           & 56.0\% (37--73) & 27.3\% (13--48) \\
    $A$                   & 0.344       & 0.140     \\
    $AR$                  & 0.317       & 0.100     \\
    \bottomrule
  \end{tabular}
\end{table}

\paragraph{Findings} Both symbols we use are the modal proposals. For
highlight, 14 of 25 participants drew a four-pointed star or sparkle
(56.0\%), and agreement falls in the high band ($AR = 0.317$). For
shadow, the crescent moon is the most frequent proposal, 6 of 22 (27.3\%),
and agreement sits on the boundary between the low and medium bands
($AR = 0.100$). The moon's lead over its runner-up, the circle
(6 against 4), is too small to be reliable at this sample size. We adopt the moon because it is the modal proposal, and because a
circle is already common in garment drawings, as buttons, eyelets and
dots.

\paragraph{Differences between categories} The two categories
differ in more than agreement. No participant answered the highlight
item with hatching, while 6 of 28 did so for shadow, under the same
instruction on the same sheet (paired, exact McNemar test \cite{mcnemar1947}, $p = 0.031$). The
preceding item listed hatching as one way to depict shadow, and 10
participants ticked it, but only 2 of those 10 were among the 6 who
gave hatching as their symbol.

\paragraph{Scope} Most participants had no formal training, so the preference for the star is unlikely to come from a convention taught in class. The study is large enough to
establish the star as the highlight symbol and the difference in
hatching between the two categories, but not the moon's lead over the
circle; its other limits are discussed in Sec.~\ref{sec:limitations}.

  \subsection{Representation}
  \label{sec:representation}

\paragraph{Two symbols, two roles} A \emph{star} marks where light
falls: the surface under it faces the light and is among the brightest
parts of the garment. Like the highlight sketches of Poulin and
Fournier~\cite{poulin1992lights}, stars specify the light through its
effect, and their placement carries the
direction. A \emph{moon} marks where the garment should be in shadow, but
says nothing about where the light is. The split follows the elicitation (Sec.~\ref{sec:elicitation}):
highlight had a clear symbol and shadow did not, and a fifth of the
participants answered the shadow item with a shading technique rather
than a symbol. In the notation, therefore, the light direction is carried by the
stars, and a moon asks only that its location be in shadow; the network
itself reads both symbols together (Sec.~\ref{sec:irradiance}).

\paragraph{Moon marks and geometry} A shadow has one of three causes:
a surface turned away from the light, a shadow cast by another part of
the garment, or a crease. The first already follows from the light the
stars set, so a moon marks shadow that the light alone does not produce. We produce it by changing the surface: the system bends the garment so that the marked region falls into
shadow under the light the stars define (Sec.~\ref{sec:displacement}). Existing pipelines would instead paint the region dark in the texture.

\paragraph{Marked sketch and clean sketch} The designer draws the
garment, then stamps the marks as fixed-size icons on a layer above the
drawing. Stamping the marks fixes their shape and size, so the network reads where a mark is placed rather than how it was drawn, and keeps them on a separate layer that can be removed for the albedo. The pipeline therefore receives two images: the \emph{marked}
sketch, with the layer, and the \emph{clean} sketch, the drawing alone.
The U-Net reads the marked sketch, and the light CNN reads it through
the U-Net's bottleneck; the albedo generator sees only the clean sketch,
so the marks never enter the albedo.

\paragraph{Placement} A mark lies inside the garment, clear of the
silhouette, and a star and a moon never overlap. Marks are stamped at
one fixed size, the size used in training. An input sketch carries at least one star and at least one moon.

\begin{figure*}[!h]
  \centering
  \includegraphics[width=\textwidth]{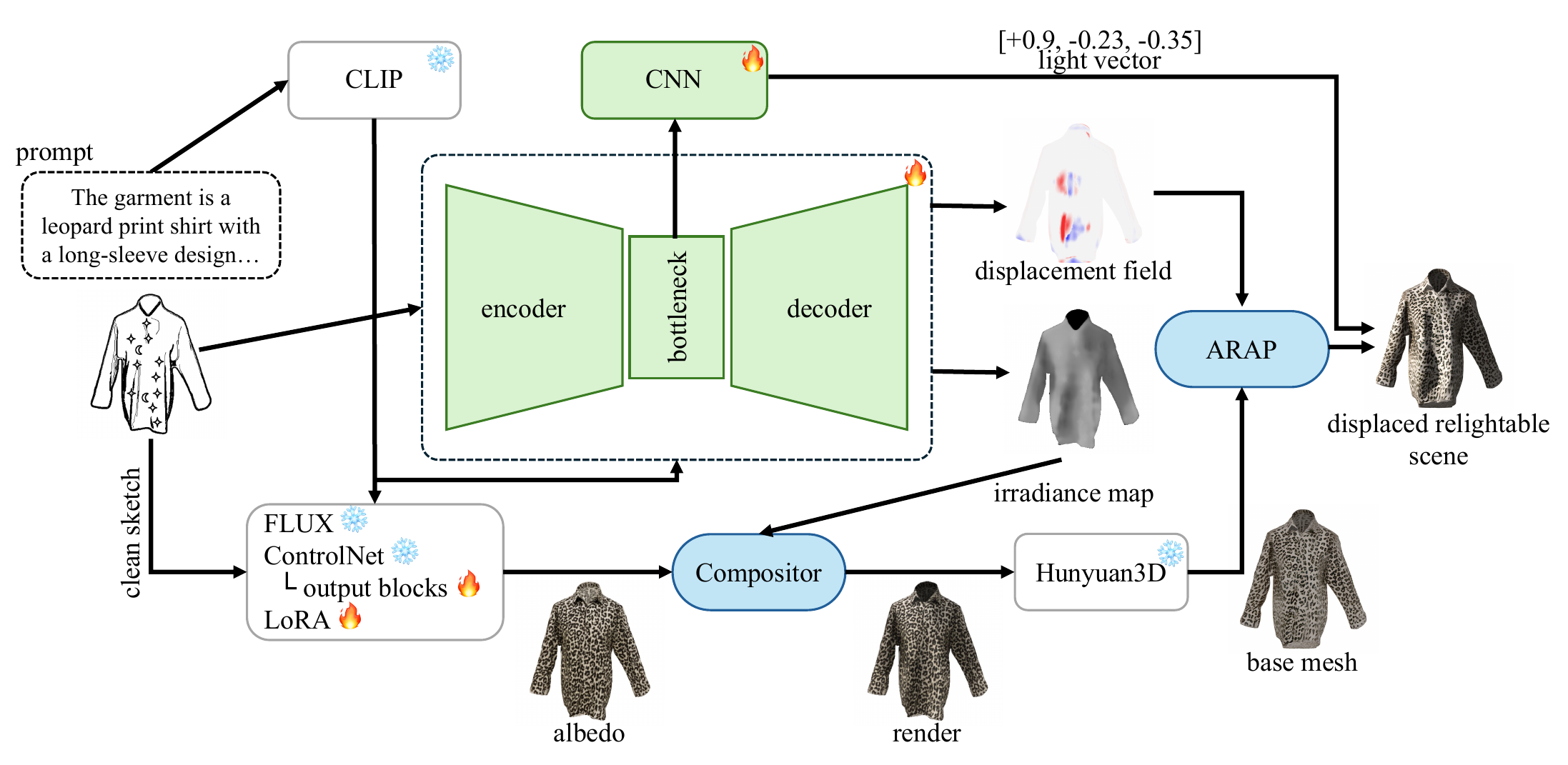}
\caption{\textbf{Pipeline overview}
From a marked sketch and a text prompt to a relightable garment
scene. The U-Net predicts an irradiance map and a field from the
marked sketch and the prompt; a light CNN reads its bottleneck for the
light direction; a FLUX ControlNet paints the albedo from the clean
sketch and the prompt. Albedo and irradiance are composited into the
image Hunyuan3D reconstructs. The predicted field together with its
crests forms the displacement field, which an as-rigid-as-possible
solve applies to that mesh. Snowflakes mark frozen weights, flames
mark trained ones; the U-Net is trained first, then frozen, and the
light CNN trained on top of it.
\emph{Prompt:} ``a leopard print shirt with a long-sleeve
design\ldots''}
\label{fig:arch}
\end{figure*}

\section{Method}
\label{sec:method}

This section describes the pipeline of Figure~\ref{fig:arch}.
Secs.~\ref{sec:irradiance}--\ref{sec:displacement} describe each stage in the order of the figure.

\subsection{Overview}
\label{sec:method-overview}

Figure~\ref{fig:arch} shows the pipeline. A U-Net reads the marked
sketch and the prompt and predicts an irradiance map and a signed
field describing folds; a light CNN reads the U-Net's bottleneck for
the light direction (Sec.~\ref{sec:irradiance},~\ref{sec:field}). A
FLUX ControlNet paints the albedo from the clean sketch and the
prompt (Sec.~\ref{sec:albedo}). Albedo and irradiance are composited
and reconstructed by Hunyuan3D-2.1 (Sec.~\ref{sec:reconstruction}),
and the predicted field is combined with crests placed using the
predicted light to form the displacement field, which is applied to
the mesh; the result is exported with the predicted light
(Sec.~\ref{sec:displacement}).

\subsection{Predicting irradiance and light direction}
\label{sec:irradiance}

The U-Net \cite{ronneberger2015unet} takes the marked sketch as its
only image input, both symbols together, and is conditioned on the
prompt through a frozen CLIP text encoder \cite{radford2021clip} by
cross-attention at the encoder, the bottleneck and the decoder. Its
first output, from a $1\times1$ convolution on the decoder, is an
irradiance map. To
build its target we render the training scene again with the garment
made white and matte, which leaves only the light that reaches each
point, shadows included.

The light direction comes from the bottleneck. A small light CNN
takes the same tensor the decoder reads and returns a unit vector for
the scene's directional light.

\subsection{Predicting where the surface folds}
\label{sec:field}

The U-Net's second output is a predicted field, a single signed value
at every point of the garment from a $1\times1$ convolution on the
decoder. Positive values push the surface out along its normal and
negative values pull it in. Together with the crests of
Sec.~\ref{sec:displacement}, it forms the displacement field that
drives the mesh.

Its target is built from the reference garment's surface normals: we
measure how fast they turn horizontally and vertically and combine
the two rates into a signed value that responds to bending rather
than to depth. The head is supervised inside a disc of radius $24$
pixels around each visible moon mark and nowhere else, except on the
training sketches we draw without moons, where it is supervised to
predict no displacement anywhere, so that it does not displace a
garment no one has marked. Because the target carries both signs, a
disc around a mark holds both the fold and the cloth beside it, and
the head learns the shape of a crease rather than only its depth. The
target is rescaled per garment, so the prediction ranks crease
strength within a garment rather than measuring an absolute distance.

One caveat on the training signal: the dataset places moon marks at
the negative extrema of this same measure, so during training the
head usually finds a crease at a mark. Sec.~\ref{sec:shading}
therefore also evaluates on marks we place ourselves, which that
measure did not choose.

\subsection{Albedo synthesis}
\label{sec:albedo}

Surface color is produced by FLUX.1 \cite{flux2024} with a ControlNet
\cite{zhang2023controlnet,shakker2024controlnet} that conditions it
on the clean sketch and the text prompt. Because a sketch carries no
color, the prompt is the only input that specifies it. What the
branch has to produce is base color without the scene's illumination,
so that the predicted irradiance can be applied afterwards rather
than being baked into the color. A pretrained image generator does not do this by default, since it is
trained to produce lit, shaded pictures. We therefore train two parts of the ControlNet and leave both it and
FLUX otherwise frozen: a rank-$64$ LoRA adapter \cite{hu2022lora} on
its attention and MLP layers, $69$M parameters, and the output layers
of its double-stream blocks, the last stage before it passes its
signal to FLUX, $47$M.

\subsection{Composition and reconstruction}
\label{sec:reconstruction}

The predicted irradiance $E$ and the albedo $\rho$, decoded from sRGB,
are combined in linear space into a shaded image
\begin{equation}
  \begin{aligned}
    C &= \rho\,(E + a) + s\,(1-r)\,f(\rho)\left(\frac{E}{E_{\max}}\right)^{n(r)},\\[6pt]
    n(r) &= \min\!\left(\max\!\left(\tfrac{2}{r^{2}} - 2,\, 1\right),\, 512\right).
  \end{aligned}
  \label{eq:composite}
\end{equation}
Both terms are restricted to the garment silhouette. The ambient term
$a = 0.15$ keeps regions the light does not reach, such as the inside
of a collar, at the color of the albedo instead of black. The second
term adds a highlight. We have no surface normals in image space, but the irradiance varies largely with the cosine of the angle to the light, so we raise $E/E_{\max}$, with $E_{\max}$ the larger of $1$ and the maximum of $E$, to a Blinn--Phong exponent \cite{blinn1977models}
derived from the roughness $r$ \cite{walter2007microfacet}. The
roughness is a constant looked up from the prompt; $(1-r)$ weakens the
highlight on rough fabric, $s = 0.25$ sets its strength, and
$f(\rho)$, a smoothstep from $0$ to $1$ as the largest sRGB channel of
$\rho$ rises from $0$ to $0.02$, suppresses it at black and near-black
albedo, so that silhouette pixels where the albedo branch painted no
garment are not turned grey; a garment painted pure black loses its
highlight as well. When more than a quarter of the garment receives almost no
predicted irradiance, which happens on some hand-drawn sketches, we
first raise the garment to a floor of $0.35$ of the median of the lit
pixels, so that the image is not mostly black. The result is
normalised by its peak and gamma-encoded.

The silhouette becomes the alpha channel of this image, and the image
is passed to Hunyuan3D-2.1 \cite{hunyuan2025hunyuan3d21}, which builds
the shape and paints the texture from it. The composite does not set
the lighting of the output, which is carried by the directional light
(Sec.~\ref{sec:displacement}); it gives Hunyuan3D the kind of input it
is built for, a lit image rather than flat color. Its painting stage is
designed to remove that illumination from the albedo it paints
(Sec.~\ref{sec:related-text-image-3d}); where the removal is
incomplete, some of the composited shading stays in the painted
texture. The mesh it returns is a thin shell, an outer wall facing the
camera with an inner wall just behind it, which the displacement stage
must respect (Sec.~\ref{sec:displacement}).

\begin{figure*}[!h]
  \centering
  \includegraphics[width=\textwidth]{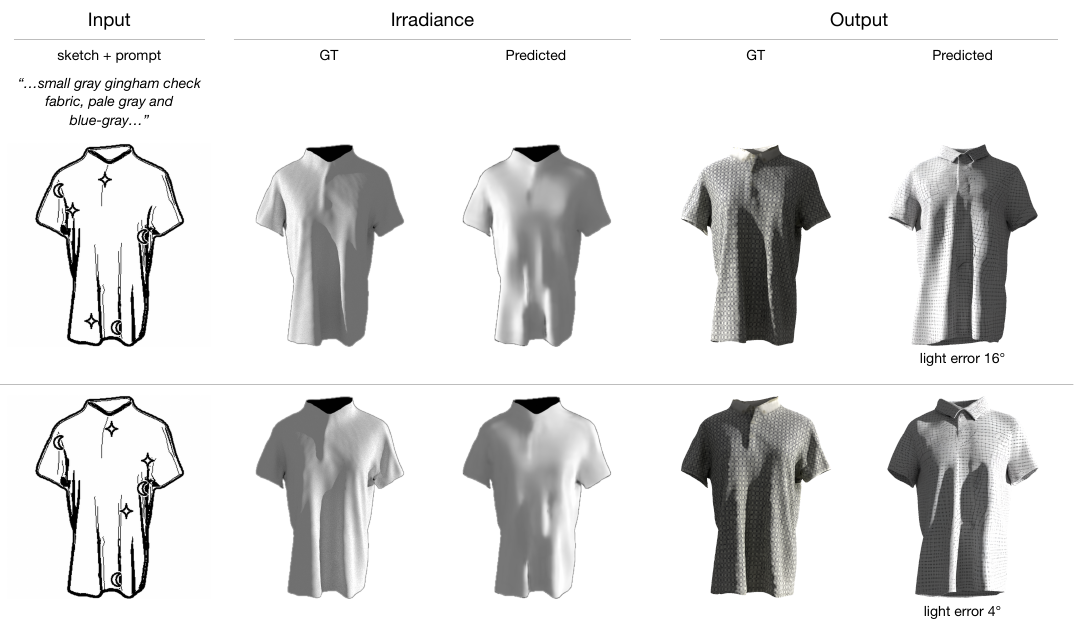}
\caption{\textbf{Light prediction}
The same garment is marked twice: the moon marks stay in the same place, and
only the star marks move.
\emph{Input:} the marked sketch and the text prompt.
\emph{Irradiance:} the ground-truth irradiance map of the reference garment
under the true light (GT), and the irradiance map predicted by our network.
\emph{Output:} the reference garment under the true light (GT), and our
displaced, textured mesh under the predicted light. The number under each
output is the angle between the predicted and the true light direction.
Moving the star marks moves the predicted light to the other side of the garment.
The prompt is the same in both rows, so the predicted albedo is identical.}
\label{fig:light-prediction}
\end{figure*}

\subsection{Displacement and scene assembly}
\label{sec:displacement}

The goal of the displacement is to put each moon-marked region into
shadow under the predicted light. After adding the crests described
below, we use the resulting field to request movement at mesh
vertices. At vertex $i$, the field value $d_i$ requests motion along
the vertex normal, outward if positive and inward if negative, and a
larger magnitude requests more movement. The combined field is
normalised by its own 95th-percentile magnitude on that garment and
scaled to the size of the mesh; inward requests are then capped at a
fraction of the local shell thickness, so that no dent asks the outer
wall to pass through the inner one. The same value sets the weight of
the request, $w_i = \lambda\,(|d_i|/\max|d|)^{\gamma}$, where
$\gamma = 0.5$ and $\lambda$ balances the requests against rigidity.
These weighted requests are soft positional constraints on
camera-facing vertices. Vertices facing away from the camera are
anchored to their original positions. An as-rigid-as-possible solve
\cite{sorkine2007arap}, using the spokes-and-rims energy of
Chao et al.~\cite{chao2010}, computes the final mesh by following the
requests while preserving the surrounding surface. The solve and
subsequent smoothing can still carry a few vertices through the shell
(Sec.~\ref{sec:limitations}).

\begin{figure*}[!ht]
  \centering
  \includegraphics[width=\textwidth]{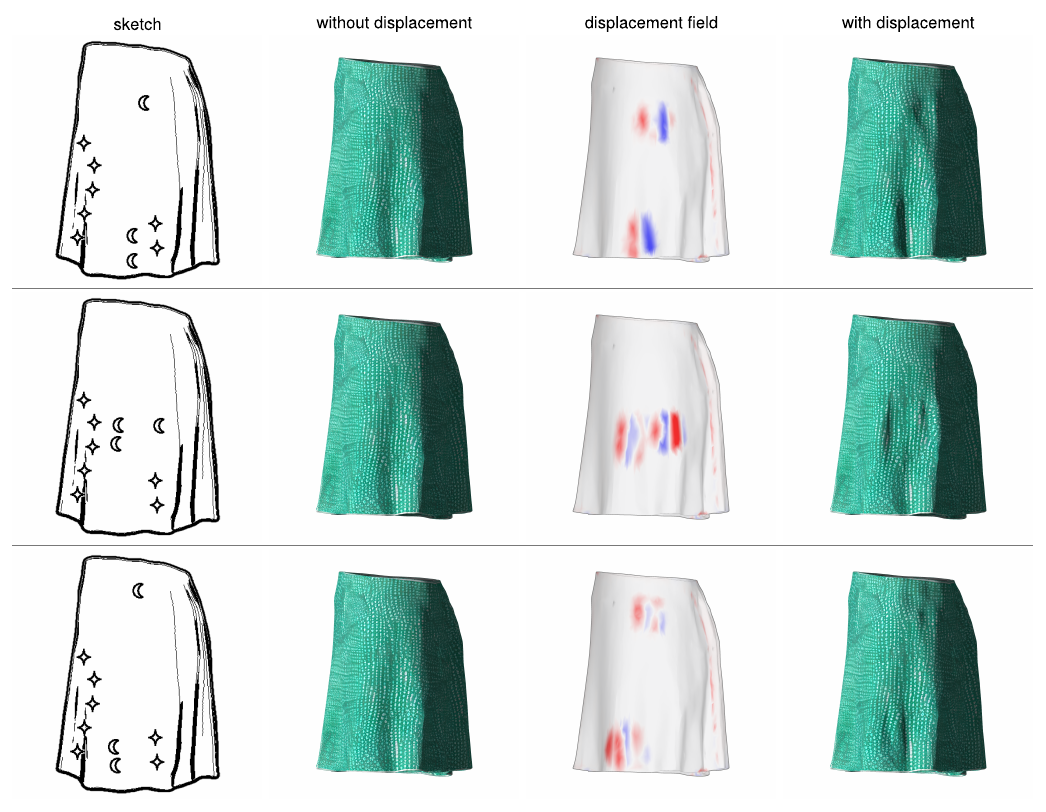}
\caption{\textbf{Geometry displacement}
Each row is the same garment with the moon marks in a different
place; the star marks stay fixed.
\emph{sketch}: the marked input.
\emph{without displacement}: the reconstruction our displacement
stage starts from.
\emph{displacement field}: what drives the solve, the predicted field
together with its crests, red outward and blue inward.
\emph{with displacement}: the same mesh after it is applied.
One reconstruction is reused across all three rows, so every
difference comes from the displacement stage.}
  \label{fig:displacement}
\end{figure*}

The predicted field decides where the surface folds, which way and
how strongly. But the field head is trained on crease shape alone
(Sec.~\ref{sec:field}), so a fold it predicts may fall on either side
of the mark, and only one side casts a shadow onto it under the
predicted light. To secure that side, we place one bump per mark
alongside the predicted field, which we call the crest: beside each
moon mark, on the side facing the predicted light, the surface is
raised so that the mark lies on its shadowed side. The crest's size
is set against that garment's own predicted field, through the 95th
percentile of its magnitude, so it sits in proportion to the folds
the network predicted there. It secures a rise rather than imposing
one: inside its footprint we keep whichever is larger, the crest or
the predicted field. The predicted light sets which side of the mark
the crest sits on; its footprint and its distance from the mark are
set by a rule.
Figure~\ref{fig:displacement} shows the result for different
placements of the moon marks on the same garment.

The displaced mesh is exported as a glTF scene
\cite{khronos2018lights} with the painted base-color texture, a
single roughness value selected from the prompt
(Sec.~\ref{sec:reconstruction}), and the predicted directional light
as an explicit directional source.

\section{Results}

\begin{figure*}[!ht]
  \centering
  \includegraphics[width=\textwidth]{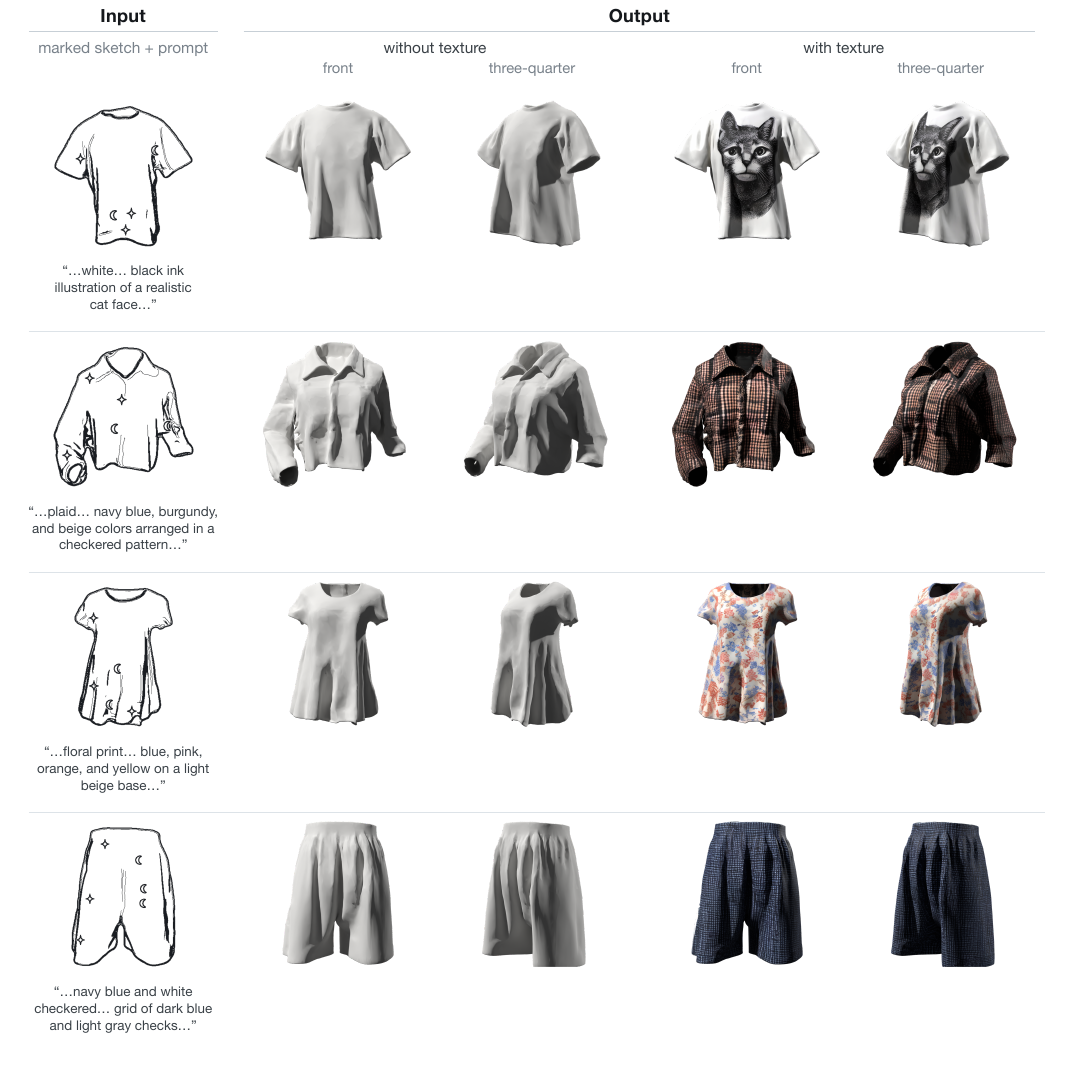}
\caption{\textbf{End-to-end results}
The light and the shading follow the marks. Each row shows one garment. \emph{Input:} the marked sketch and an excerpt of
the text prompt. \emph{Output:} the displaced mesh under the
directional light predicted from its marks, first without texture and
then textured, each from the front and at a three-quarter view. \emph{Without texture} columns show the shading most clearly; with
texture, prints such as the plaid and the floral pattern hide much of
the shading the displacement adds.}
\label{fig:results}
\end{figure*}

We evaluate the two things the marks control: the light and the
shading of the displaced garment. Figure~\ref{fig:results} shows
outputs on four garments.

\subsection{Setup}
\label{sec:setup}

We evaluate on DeepFashion3D \cite{zhu2020deepfashion3d}, split at
the garment level so that no garment appears in more than one part,
stratified by garment type: $968$ garments for training, $122$ for
validation and $122$ held out. Each garment is rendered with
Mitsuba~3 \cite{jakob2022mitsuba3} under each of three lights drawn uniformly by area from the
directions that face the camera, with elevations above
$-15^{\circ}$, giving $366$ held-out renders. Each render gets its own marked sketch, within the
placement constraints of Sec.~\ref{sec:representation}: star marks
where that render's ground-truth irradiance is brightest, and moon
marks where the reference garment creases most sharply, by the
measure of Sec.~\ref{sec:field}. A sketch carries about nine moon
marks and eight star marks.

The two claims need different data. The light direction is predicted
before any mesh exists, so it is measured on all $366$ held-out
renders. Displacement needs a reconstructed mesh, which we build for
all of them, and two checks remove the ones that came out wrong: one
rejects a garment seen from the wrong side, the other a silhouette
that departs too far from the sketch it was conditioned on. Together
they remove $26$, leaving $340$ renders over $121$ garments.

Placing the moon marks at the reference garment's own creases has a
consequence for evaluation: the reconstruction may already have a
fold where a mark sits, and the displacement has little left to add
there. To measure it where nothing helps it, we annotate $112$
renders ourselves, placing the moon marks on flat cloth where the
undisplaced reconstruction is slightly brighter than the garment's
average, so nothing there is shaded yet. We keep the reconstructions
those renders already have, so this measures the displacement stage
alone rather than the whole pipeline. On these renders we also check
whether a mark ends up darker than the garment's average, scored on a
simpler shading estimate computed directly from the mesh; because
that estimate differs slightly from the render the marks were placed
on, a few marks already fall below the average before displacement.

Shading at a mark is the mean brightness inside a disc at the mark,
measured on a render of the mesh under the predicted light.

\paragraph{Implementation details}
The U-Net is trained for $60$ epochs with Adam \cite{kingma2015adam}
at a learning rate of $3\times10^{-4}$ and batch size $8$ on
$512\times512$ sketches, taking about $26$ GPU-hours on one A40. The
light CNN is then fitted for $60$ epochs on the frozen U-Net, $50$
minutes on the same hardware, selecting the epoch with the best
validation angle. The FLUX branch (Sec.~\ref{sec:albedo}) is trained
for $2500$ steps at $1\times10^{-4}$ with an effective batch of $4$,
just under four hours. One garment takes about seven minutes end to
end on a single A40: $3.3$ minutes in Hunyuan3D-2.1, $2.4$ minutes
in the albedo branch, and under a minute in everything else.
\subsection{Lighting recovery}
\label{sec:light}

On the $366$ held-out renders the light CNN recovers the light
direction to $22.9^{\circ}$ mean and $20.0^{\circ}$ median angular
error, against $47.5^{\circ}$ for the best fixed direction fitted on
the training split. The result does not depend on which
reconstructions succeeded: restricted to the $340$ renders of
Sec.~\ref{sec:setup}, the mean error and the baseline are unchanged
to one decimal. Figure~\ref{fig:light-prediction} shows one garment
annotated with the star marks on either side.

\subsection{Mark-driven shading}
\label{sec:shading}

\paragraph{Darkening at the marks}
We use the change in brightness inside each moon mark's disc as our
measure of the intended shadow. After displacement, $79\%$ of the moon
marks are darker than before, against $44\%$ of equally sized discs
placed away from any mark on the same garments ($3020$ scored discs
and $68\,000$ control discs). The control rate sits near half because
displacing a mesh shifts unmarked cloth a little in both directions,
so whether such a disc ends up lighter or darker is close to
arbitrary. Requiring a brightness decrease of at least eight grayscale levels,
the marks darken in $51\%$ of cases against $13.5\%$ of control discs. We
also run the displacement with the crest disabled: the predicted
field alone darkens $61\%$ of the marks against $47\%$ of the control
discs for that arm.

\paragraph{Marks the reconstruction does not already satisfy}
On the $112$ renders we annotate ourselves (Sec.~\ref{sec:setup}),
the displacement has no existing fold to deepen. After displacement
$86\%$ of these moon marks are darker than they were, against $45\%$
of control discs on the same garments, and with the crest disabled
the field alone darkens $57\%$ against $47\%$. Against the garment's
average, $9\%$ of marks are darker before displacement and $73\%$
after.

\section{Limitations}
\label{sec:limitations}

Our system has limitations in how far a mark can be driven, in how
the crest is placed, in the artifacts the displacement can leave, in
the scene and material model, and in what we have evaluated. We
discuss each in turn.

\paragraph{Why some marks do not darken}
Two things limit how far a mark can be driven. The field head learns
from the DeepFashion3D garment, but the field it predicts is applied
to Hunyuan3D's reconstruction, which shares a silhouette but not a
surface. And inward motion is capped at a fraction of the shell's
wall thickness; the cap binds on $314$ of the $340$ evaluated renders, and
where it binds the requested depth is typically several times what
the shell allows.

\paragraph{Crest placement}
Beside each mark we add one crest, and its footprint and its distance
from the mark are set by a rule rather than predicted from the
sketch. The rest of the displaced surface comes from the predicted
field.

\paragraph{Artifacts}
On $100$ of the $340$ held-out renders the displacement leaves a
visible artifact, most often a small blob-like distortion, or the
outer wall passing through the inner one. We exclude those renders
from our figures, though not from the measurements above.

\paragraph{Scene and material assumptions}
We assume a single directional light, with no colored or area
sources, no translucency and no interreflection. Roughness is looked
up from the material named in the prompt, and
is constant over a garment. Where Hunyuan3D does not fully remove the
shading we composite into its input, that shading stays in the texture
(Sec.~\ref{sec:reconstruction}) and does not follow the light. The
displacement is shaped for the predicted directional light, so moving the light afterwards keeps the folds but not necessarily the shadow each fold casts onto its mark.

\paragraph{Coverage}
The elicitation study involved art students from a single
institution, so we cannot say whether professional designers would
choose the same symbols. Our quantitative evaluation uses synthetic
sketches rendered from DeepFashion3D, with marks placed by rule or by
us rather than drawn by hand; hand-drawn sketches, such as the one in
Figure~\ref{fig:teaser}, appear only in qualitative results. We have
also not tested whether people read the symbols as intended, or how
far a mark can be moved before the result changes.

\section{Conclusion}
\label{sec:conclusion}

We have presented a system that reads the star and moon marks a designer places on a garment sketch and returns a 3D scene in which the light is an explicit, editable source
and the shadows the marks ask for are made by displacing the surface. Two marks, a
star where light falls and a moon where the garment should be in
shadow, came out of an elicitation study with art students and are
stamped on the sketch itself. From the marked sketch and a text
prompt, the network predicts the light's direction and where the
fabric should fold, recovering the light direction to $22.9^{\circ}$
against $47.5^{\circ}$ for the best fixed direction fitted on the
training split. Displacing the surface darkens $79\%$ of the moon
marks on held-out renders, and $86\%$ on renders we mark ourselves on
flat cloth, where the reconstruction offers no fold to deepen.

One part remains open. The crest beside each mark takes its height
from the predicted field and its side from the predicted light, but
its footprint and its distance from the mark are fixed by a rule.
Predicting those is our next step.

\section*{Acknowledgments}
We thank Dr. İrem Şot for her help in preparing the elicitation study and
for hosting the information sessions in her class, and the students who
took part.

\bibliographystyle{alpha}
\bibliography{main}
\end{document}